\documentclass[prl,aps,twocolumn]{revtex4}

\newcommand{\plb}[2]{{\em Phys. Lett.}              {\bf #1B}, #2 }
\newcommand{\npb}[2]{{\em Nucl. Phys.}              {\bf B#1}, #2 }
\newcommand{\npp}[2]{{\em Nucl. Phys. Proc. Suppl.} {\bf  #1}, #2 }
\newcommand{\pr }[2]{{\em Phys. Rep.}               {\bf  #1}, #2 }
\newcommand{\prt}[2]{{\em Phys. Rev.}               {\bf D#1}, #2 }
\newcommand{\pru}[2]{{\em Phys. Rev. Lett.}         {\bf  #1}, #2 }
\newcommand{\zpc}[2]{{\em Z. Phys.}                 {\bf C#1}, #2 }

\newcommand{\epc}[2]{{\em Eur. Phys. J.}            {\bf C#1}, #2 }
\newcommand{\ijm}[2]{{\em Int. J. Mod. Phys.}       {\bf A#1}, #2 }
\newcommand{\con}[2]{                               {\bf  #1}, #2 }
\newcommand{\etal}{{\em et al.}}
\newcommand{\ibid}{{\em ibid.}}

\newcommand{\aspi}{{\hat\alpha_s\over \pi}}
\newcommand{\aspis}{{\hat\alpha_s^2 \over \pi^2}}
\newcommand{\five}{\hspace{5pt}}

\newcommand{\be}{\begin{equation}}
\newcommand{\ee}{\end{equation}}
\newcommand{\ba}{\begin{array}}
\newcommand{\ea}{\end{array}}

\def\msbar{\mbox{{\footnotesize{$\overline{\rm MS}$}}} }

\newcommand{\lsim}{\buildrel < \over {_\sim}}

\def\al2{\frac{\alpha^2}{\pi^2}}

\begin{document}

\title{
Precision Determination of Heavy Quark Masses
and the Strong Coupling Constant}
\author{Jens Erler$^{a,b}$ and Mingxing Luo$^{a,c}$}
\address{
$^{a}$Dept.\ of Physics and Astronomy, University of Pennsylvania,
209 S.\ 33rd St., Philadelphia, PA 19104-6396, USA \\
$^{b}$Instituto de F\'\i sica, Universidad Nacional Aut\'onoma de M\'exico, 
Apdo.\ Postal 20-364, 01000 M\'exico D.F., M\'exico \\
$^{c}$Zhejiang Institute of Modern Physics, Dept.\ of Physics,
Zhejiang University, Hangzhou, Zhejiang 310027, P.R. China}

\date{July 2002}

\begin{abstract}
We present a new QCD sum rule with high sensitivity to the continuum regions of
charm and bottom quark pair production.  Combining this sum rule with existing 
ones yields very stable results for the \msbar quark masses, 
$\hat{m}_c (\hat{m}_c)$ and $\hat{m}_b (\hat{m}_b)$.  We introduce 
a phenomenological parametrization of the continuum interpolating smoothly 
between the pseudoscalar threshold and asymptotic quark regions.  Comparison of
our approach with recent BES data allows for a robust theoretical error 
estimate. The parametric uncertainty due to  $\alpha_s$ is reduced by 
performing a simultaneous fit to the most precise sum rules and other high 
precision observables.  This includes a new evaluation of the lifetime of 
the $\tau$ lepton, $\tau_\tau$, serving as a strong constraint on $\alpha_s$.  
Our results are
$\hat{m}_c (\hat{m}_c)  = 1.289^{+0.040}_{-0.045}$~GeV,
$\hat{m}_b (\hat{m}_b)  = 4.207^{+0.030}_{-0.031}$~GeV 
(with a correlation of 29\%), and
$\alpha_s (M_Z) [\tau_\tau] = 0.1221_{-0.0023}^{+0.0026}$.
\end{abstract}


\maketitle

The determination of the fundamental Standard Model (SM) parameters is 
important not only in its own right, but also as a SM test when results from 
various sources are compared.  Such comparisons can foster our understanding 
of SM dynamics (such as strong QCD effects) or may ultimately lead to hints 
of new physics beyond the SM.  Moreover, precise values of the SM parameters 
can be compared against the predictions of more fundamental theories.  
Well known examples are Grand Unified Theories~\cite{Langacker:1980js} which 
typically predict values for the strong coupling constant, $\alpha_s$, and 
the mass ratio $m_b/m_\tau$.

It is generally difficult to obtain reliable information on quark masses.  
The Particle Data Group~\cite{Hagiwara:2002} lists only ranges for their 
values, indicating a lack of confidence in the theoretical methods used 
to evaluate them.  Indeed, $\alpha_s$ is quite large at the mass scales of 
the bottom and charm quarks, questioning the convergence of perturbative QCD 
(PQCD).  Furthermore, non-perturbative (power suppressed) effects governed by 
the scale $\Lambda_{\rm QCD} \sim 0.5$~GeV could be large, thus compromising 
reliable calculations.  Two types of conditions are known to improve 
the situation:  high energy or inclusiveness.  As an example for the former, 
$\alpha_s$ and $m_b$ can be determined at LEP energies using PQCD. This yields
$\alpha_s(M_Z) = 0.1200\pm 0.0028$~\cite{Erler:sa} with very little theoretical
uncertainty, but $b$ quark effects are small and 
$\hat{m}_b(M_Z)= 2.67\pm 0.50$~GeV~\cite{Abreu:1997ey} is not well constrained.

In this Letter, we compute $\alpha_s$ from $\tau_\tau$, by definition 
an inclusive quantity and known to be quite insensitive to effects from 
non-perturbative QCD (NPQCD)~\cite{Braaten:1991qm}.  Likewise, we use
a set of inclusive QCD sum rules to derive values for $\hat{m}_c (\hat{m}_c)$ 
and $\hat{m}_b (\hat{m}_b)$.  One of these sum rules is new, and its use 
together with existing ones~\cite{Novikov:et,Shifman:bx} proves to be 
a powerful tool to constrain the continuum region of quark pair production.  
This will be particularly helpful for the case of the $b$ quark
for which precise measurements of $R(s)$ (the inclusive hadronic cross section 
normalized to the leptonic point cross section) or of $R_b(s)$ 
(exclusive cross section for $b\bar{b}$ pairs) are unavailable.

On the basis of an unsubtracted dispersion relation (UDR) it was shown 
in Ref.~\cite{Erler:1998sy} that knowledge of $m_c$, $m_b$, and $\alpha_s$ 
is sufficient to compute the charm and bottom quark contributions to 
the QED coupling $\alpha (\sqrt{t} = M_Z)$, a vital parameter entering 
the analysis of the very high precision LEP~1 and SLC data. Or conversely, 
comparison of this UDR with the more traditional approach using 
a subtracted dispersion relation (SDR) offers information on $m_c$ and 
$m_b$.  The resulting equation relates an inclusive integrated cross 
section to a difference of vacuum polarization tensors, {\it viz.}
\be
   12\pi^2 \left[ \hat\Pi_q (0) - \hat\Pi_q (-t) \right]=
   t \int_{4 m_q^2}^\infty {{\rm d} s\over s} {R_q(s)\over s + t}.
\label{sumrulet}
\ee
Eq.~(\ref{sumrulet}) defines a continuous set of sum rules parametrized 
by $t$, where the limit $t \rightarrow 0$ coincides with the first 
moment of $\Pi_q (t)$. Similarly, there is a sum rule,
\be
   \left.{12\pi^2\over n !} {d^n\over d t^n} \Pi_q(t) \right|_{t=0}
   = \int_{4 m_q^2}^\infty {{\rm d} s\over s^{n+1}} R_q(s),
\label{sumrulen}
\ee
for each higher moment, ${\cal M}_n$, as 
well~\cite{Novikov:et,Shifman:bx,Voloshin:1995sf,Eidemuller:2000rc,Kuhn:2001dm}. We now take the opposite limit in Eq.~(\ref{sumrulet}), $t\rightarrow\infty$,
and regularize the divergent expression (which will render ${\cal M}_0 < 0$ !),
\be
   {R_q (s)\over 3 Q_q^2} \longrightarrow {R_q (s)\over 3 Q_q^2} - 
   \lambda^q_1 (s) \equiv {R_q (s)\over 3 Q_q^2} - 1 -  
   {\alpha_s (\sqrt s) \over \pi}
\label{regular}
\ee
\vspace{-12pt}
$$   - \left[ {\alpha_s (\sqrt s) \over \pi} \right]^2
       \left[ {365\over 24} - 11 \zeta(3)
            + n_q \left( {2\over 3} \zeta(3) - {11\over 12} \right) \right].
$$
$Q_q$ and $n_q$ are the quark charge and the number of active flavors.  
Using expressions derived in Refs.~\cite{Chetyrkin:1996cf,Chetyrkin:1997un}, 
we can now explicitly write down the sum 
rule~(\ref{sumrulet}) for $t \rightarrow \infty$:
$$
   \sum\limits_{\rm resonances} {3\pi\Gamma^e_R\over Q_q^2 M_R 
   \hat\alpha^2 (M_R)} + \int\limits_{4 M^2}^\infty {{\rm d} s\over s}  
   {R_q^{\rm cont}\over 3 Q_q^2} - \int\limits_{\hat{m}_q^2}^\infty 
   {{\rm d} s\over s} \lambda^q_1 (s) 
$$
\vspace{-12pt}
$$
   = - {5\over 3} + \aspi \left[ 4 \zeta(3) - {7\over 2} \right] +
   \aspis  \left[ {11\over 4} \zeta(2) + {2429\over 48} \zeta(3) - \right. 
$$
\vspace{-12pt}
\be
   \left. {25\over 3} \zeta(5) - {2543\over 48} + n_q \left( {677\over 216} -
   {\zeta(2)\over 6} - {19\over 9} \zeta(3) \right) \right].
\label{sumrule0}
\ee
Here, $M_R$ and $\Gamma^e_R$ are the mass and the electronic partial width of 
resonance $R$, and $R_q^{\rm cont}$ denote the continuum regions integrated 
from $M = M_{B^\pm}$ for $b$ and $M = M_{D^0}$ for $c$.  The $\zeta(2)$ terms 
arose from the regularization~(\ref{regular}) which together with the scale 
choices $\hat{m}_q = \hat{m}_q (\hat{m}_q)$ and 
$\hat\alpha_s = \hat\alpha_s (\hat{m}_q)$ eliminates (resums) all logarithmic 
terms in Eq.~(\ref{sumrule0}). Unlike in any of the sum 
rules~(\ref{sumrulen}), $R_q^{\rm cont}$ appears unsuppressed in 
Eq.~(\ref{sumrule0}) so that $\hat{m}_q$ varies exponentially with 
the experimental information on the resonances.  An optimal approach to compute
$\alpha (M_Z)$ would first identify the sum rule most sensitive to $m_q$, and 
then use the value so obtained in theoretical expressions such as the one 
presented in Ref.~\cite{Erler:1998sy}. We will use  Eq.~(\ref{sumrule0}) 
to constrain the continuum region and work with the following {\em ansatz\/}:
$$
   {R_q^{\rm cont} (s)\over 3 Q^2_q} = \lambda^q_1 (s) 
   \sqrt{1 - {4\, \hat{m}_q^2 (2 M) \over s^\prime}} 
   \left[ 1 + \lambda^q_3 {2\, \hat{m}_q^2(2 M) \over s^\prime} \right] \approx
$$	
\vspace{-12pt}
\be
   \lambda^q_1 (4 M^2) \sqrt{1 - {4\, \hat{m}_q^2 \over s^\prime}} \left[ 1 + 
   \lambda^q_3 {2\, \hat{m}_q^2 \over s^\prime} \right] - \aspi
   {\lambda_2^q (s)\over 1 + \lambda_2^q(s)},
\label{ansatz}
\ee
where now $\hat\alpha_s = \hat\alpha_s (2 M)$, 
$s^\prime \equiv s + 4 (\hat{m}_q^2 (2 M) - M^2)$, and
$$
  \lambda_2^q(s) = \aspi \beta_0 \ln {s\over 4 M^2} = {\hat\alpha_s (2 M)\over 
  \pi} \left( {11\over 4} - {n_q\over 6} \right) \ln {s\over 4 M^2}.
$$ 
We will use the form in the second line (applying it to all moments) of 
Eq.~(\ref{ansatz}) with the corresponding change in the regularization in 
Eq.~(\ref{sumrule0}). This keeps only the leading logarithms resummed but 
allows for an analytical integration.  
Eq.~(\ref{ansatz}) coincides asymptotically with the predictions 
of PQCD for massless quarks and interpolates smoothly between the vanishing 
phase space at the pseudoscalar threshold and the strong onset of fermion pair 
production. The quark parton model predicts $\lambda^q_3 = 1$, while from third
order massive QCD corrections~\cite{Chetyrkin:1996ia} one expects 
$\lambda^q_3 > 1$ (in agreement with our results). But unlike when PQCD is 
applied to $R(s)$ directly and relatively close to the resonance region, we 
minimize the exposure to local quark-hadron duality violations by using QCD 
{\em inclusively\/} and by merely requiring stable results across the moments. 
No claim is being made about the {\em local shape\/} of $R_q$ --- we only need 
theoretical information about {\em global averages\/}.

We use the narrow resonance data~\cite{Hagiwara:2002} listed in 
Table~\ref{resonances} as the only experimental input.  The wider resonances in
the continuum region are assumed to be accounted for by our 
{\em ansatz\/}~(\ref{ansatz}) because (i) they decay almost exclusively into 
flavored hadrons; (ii) they interfere with the non-resonating part of 
\begin{table}[h]
\begin{tabular}{|c|r|r||c|r|r|}
$R$ & $M_R$ [GeV] & $\Gamma_R^e$ [keV] & 
$R$ & $M_R$ [GeV] & $\Gamma_R^e$ [keV] \\
\hline
  $J/\Psi$ & 3.09687 & 5.26     (37) & $\Upsilon(1S)$ &  9.46030 & 1.320 (50)\\
$\Psi(2S)$ & 3.68596 & 2.19     (15) & $\Upsilon(2S)$ & 10.02326 & 0.520 (32)\\
           &         &               & $\Upsilon(3S)$ & 10.35520 & 0.476 (78)\\
\end{tabular}
\caption[]{Resonance data~\cite{Hagiwara:2002} used in the analysis. 
The uncertainties from the resonance masses are negligible.}
\label{resonances}
\end{table}
the continuum rendering a common treatment virtually impossible; 
(iii) the $\delta$-function approximation (which is perfect for the narrow
resonances) becomes successively worse; (iv) the philosophy of our 
{\em ansatz\/} supposes that it averages over local cross-section fluctuations;
and (v) we wish to compare Eq.~(\ref{ansatz}) directly to experimental data on
the charm continuum region such as from Beijing~\cite{Bai:2001ct}.
\begin{table}[bth]
\begin{tabular}{|c|c|r|r|r|}
$n$ & resonances & continuum & total & theory \hspace{8pt} \\
\hline
0 & 1.16 (6) &$-3.03 \pm 0.37$ &$-1.86 \pm 0.37$ & {\bf input} (\five 4) \\
1 & 1.12 (6) & $1.04 \pm 0.14$ & $2.16 \pm 0.16$ & 2.19        (\five 6) \\
2 & 1.10 (7) & $0.37 \pm 0.07$ & $1.47 \pm 0.10$ & 1.49        (\five 9) \\
3 & 1.10 (7) & $0.17 \pm 0.04$ & $1.27 \pm 0.08$ & 1.26             (14) \\
4 & 1.11 (7) & $0.09 \pm 0.02$ & $1.20 \pm 0.08$ & 1.16             (20) \\
5 & 1.13 (7) & $0.05 \pm 0.01$ & $1.18 \pm 0.08$ & 1.10             (31) \\
\hline
0 & 1.17 (5) &$-52.44\pm 1.24$ &$-51.27\pm 1.24$ & {\bf input} (\five 2) \\
1 & 1.24 (5) & $3.12 \pm 0.53$ & $4.36 \pm 0.54$ & 4.51        (\five 2) \\
2 & 1.31 (5) & $1.33 \pm 0.30$ & $2.64 \pm 0.31$ & 2.79        (\five 3) \\
3 & 1.40 (5) & $0.75 \pm 0.19$ & $2.15 \pm 0.20$ & 2.27        (\five 5) \\
4 & 1.50 (5) & $0.48 \pm 0.13$ & $1.98 \pm 0.14$ & 2.06        (\five 7) \\
5 & 1.61 (5) & $0.33 \pm 0.10$ & $1.94 \pm 0.11$ & 1.99             (10) \\
6 & 1.74 (6) & $0.23 \pm 0.07$ & $1.98 \pm 0.09$ & 1.98             (14) \\
7 & 1.89 (6) & $0.17 \pm 0.05$ & $2.06 \pm 0.08$ & 2.03             (19) \\
\end{tabular} 
\caption[]{Results for the lowest moments, ${\cal M}_n$, defined in 
Eq.~(\ref{sumrulet}) for $n = 0$ ($t \rightarrow \infty$) and 
Eq.~(\ref{sumrulen}) for $n \geq 1$.  The upper (lower) half of the Table 
corresponds to the charm (bottom) quark.  Each moment has been multiplied by 
$10^n\mbox{GeV}^{2n}$ ($10^{2n+1}\mbox{GeV}^{2n}$). The continuum error is 
from $\Delta\lambda_3^{b,c} = \pm 1.47$. The last column shows the theoretical 
prediction for $\hat{m}_c (\hat{m}_c) = 1.289$~GeV, 
$\hat{m}_b (\hat{m}_b) = 4.207$~GeV, and $\alpha_s (M_Z) = 0.1211$, where 
the uncertainty is our estimate for the truncation error (see text).}
\label{moments}
\end{table}
The narrow resonance  contribution to the various moments is shown in 
the second column of Table~\ref{moments}. The $\Gamma^e_R$ are obtained from 
constrained fits~\cite{Hagiwara:2002} to a great number of measurements 
independently for each resonance, and should have very small correlations. 
We will therefore combine their propagated errors in quadrature. The 3rd 
column gives the continuum contribution, and the 4th column shows the totals 
to be compared with the theoretical moments in the last column, {\em viz.}
\be
   {\cal M}_n^{\rm theory} = {9\over 4} Q_q^2 
   \left( {1\over 2\hat{m}_q(\hat{m}_q)} \right)^{2n} \bar{C}_n.
\label{theory}
\ee
The $\bar{C}_n$ are known up to ${\cal O}(\alpha_s^2)$ and from
Refs.~\cite{Chetyrkin:1996cf,Chetyrkin:1997mb} where they were computed
for arbitrary renormalization scale $\mu$.  It seems appropriate to choose 
$\mu = \hat{m}_q (\hat{m}_q)$, eliminating all logarithmic terms as there is 
only one scale in the problem.  Indeed, the authors of Ref.~\cite{Kuhn:2001dm},
who have chosen $\mu = 3$~(10)~GeV for the charm (bottom) quark and then 
evolved to $\mu = \hat{m}_q$, report a variation over the first 5~(7)~moments 
of 122~(312)~MeV. (For larger moments the $\alpha_s$ 
expansion~\cite{Broadhurst:1994qj} of the gluon condensate 
contribution~\cite{Shifman:bx} breaks down.)  Using the same 
moments~\cite{Kuhn:2001dm} but choosing $\mu = \hat{m}_q$
instead, we observe a variation of less than 27~(16)~MeV.  This impressive 
improvement clearly overcompensates for the larger $\hat\alpha_s$.
We will choose $\mu = \hat{m}_q$ in the following.  As for the theoretical
uncertainty associated with the truncation of the perturbative series, we
use the method suggested in Ref.~\cite{Erler:1999ug}.  It exploits the fact
that once stripped off all group theoretical factors, the coefficients
appearing in PQCD (to the very least for highly inclusive quantities defined 
in the Euclidean domain) are strictly of order unity\cite{Erler:1999ug}.  
Since one can easily determine the largest group theoretical factor in 
the next uncalculated perturbative order, this offers a reliable and 
{\em transparent\/} way to assess truncation errors.  In our case this 
yields the error estimate,
\be
   \pm N_C Q_q^2 C_F C_A^2 {\hat\alpha_s^3 (\hat{m}_q)\over \pi^3}
       \left( {1\over 2 \hat{m}_q(\hat{m}_q)} \right)^{2n},
\label{error}
\ee
($N_C = C_A = 4 C_F = 3$) corresponding to $\pm 16 \hat\alpha_s^3/\pi^3$ in 
the $\bar{C}_n$. Comparing the corresponding estimate against the exactly known
coefficients of the first eight moments up to order 
$\alpha_s^2$~\cite{Chetyrkin:1996cf,Chetyrkin:1997mb} shows that with
$\mu = \hat{m}_q$, 23 of 24 coefficients are within the estimate, while only 
one coefficient would have been underestimated by a factor 
$\approx 1.437$.  This seems to be a reasonable state of affairs for 
a $1\sigma$ error estimate and corresponds to $\pm 20$~MeV for 
$\hat{m}_c(\hat{m}_c)$ from ${\cal M}_1$, while variation of 
the renormalization scale~\cite{Kuhn:2001dm} assesses this error to only 1~MeV,
which is optimistic.  
We show the estimate~(\ref{error}) in the last column of Table~\ref{moments}.
\begin{table}[t]
\begin{tabular}{|c|ccc|cc|}
n & BES & $\lambda^c_3 = 0.50$ & $\lambda^c_3 = 1.97$ & BES & $\Psi(3S)$ \\
\hline
0 & 5.51      (35) & 5.50 & 7.19 & 0.215      (39) & 0.348      (54) \\
1 & 3.02      (19) & 3.01 & 3.98 & 0.151      (27) & 0.245      (38) \\
2 & 1.68      (11) & 1.68 & 2.25 & 0.106      (19) & 0.172      (27) \\
3 & 0.95 (\five 6) & 0.96 & 1.29 & 0.074      (13) & 0.121      (19) \\
4 & 0.55 (\five 4) & 0.55 & 0.76 & 0.052 (\five 9) & 0.085      (13) \\
5 & 0.32 (\five 2) & 0.33 & 0.45 & 0.037 (\five 6) & 0.060 (\five 9) \\
\end{tabular}
\caption[]{The left part shows contributions to the charm moments 
($\times 10^{n+1}\mbox{GeV}^{2n}$) from $2 M_{D^0} \leq \sqrt{s} \leq 4.8$~GeV,
and the right part from $2 M_{D^0} \leq \sqrt{s} \leq 3.83$~GeV.  Following 
Ref.~\cite{Kuhn:2001dm}, we computed the columns labeled BES by subtracting 
from the threshold data on $R(s)$ the average, $\bar{R}$, below threshold. (We
applied corrections for the leading $s$-dependence.) The errors combine 
the statistical and uncorrelated systematic ones of $\bar{R}$ with those in 
the continuum region and with the common systematics ($\leq 3.5\%$) of 
the difference.}
\label{table:BES}
\end{table}
The last two columns of that Table would agree within errors even if we had
chosen significantly smaller variations in $\lambda_3^b$ and especially 
$\lambda_3^c$ ($\Delta\lambda_3^{b,c} = \pm 1.47$ accounts for the error
introduced by our {\em ansatz\/} and is above and beyond the variations induced
by the fit parameters). The reason for our more conservative error is shown in 
Table~\ref{table:BES}.  It shows that Eq.~(\ref{ansatz}) with 
$\lambda_3^c = 0.50$ reproduces the $n$ dependence of the moments computed from
recent data by the BES Collaboration~\cite{Bai:2001ct} remarkably well. 
However, our method favors $\lambda_3^c \approx \lambda_3^b \approx 1.97$,
and thus 30 to 40\% larger contributions. Table~\ref{table:BES}
also compares the BES data to the $\Psi(3S)$ 
contribution~\cite{Hagiwara:2002} in the narrow width approximation.  Even 
assuming that the $\Psi(3S)$ resonance ($M_{\Psi(3S)} = 3.7699$~GeV) 
saturates the charm cross-section in that region, we observe a direct 
{\em experimental\/} $2\sigma$ discrepancy between Ref.~\cite{Bai:2001ct} 
and $\Gamma^e_{\Psi(3S)} = 0.26\pm 0.04$~keV~\cite{Hagiwara:2002}.
Thus there is a discrepency between the theory and the BES data, though the theory
does seem to be consistent with the $\Psi(3S)$ data.
This constitutes a great puzzle which needs to be resolved in the future.
We may be able to quote smaller errors after this situation has been resolved.
Nevertheless, the quark masses can still be determined precisely through the sum rule approach.

There is a possible contribution from the gluon condensate~\cite{Shifman:bx}.
It is known up to ${\cal O}(\alpha_s)$~\cite{Broadhurst:1994qj}, but its actual
value is not well known.  Its inclusion lowers the extracted quark masses, 
increases $\lambda_3^c$, and sharpens the discrepancy with the BES data.
We can bound its value to $\lsim 0.07\mbox{ GeV}^4$ by demanding $n$ 
independent results within the uncertainties. We use this bound (with a central
value of zero) to account collectively for non-perturbative uncertainties. 
They induce errors of about 29~MeV into $\hat{m}_c(\hat{m}_c)$ ($n = 2$) and 
2.4~MeV into $\hat{m}_b(\hat{m}_b)$ ($n = 6$).

The parametric uncertainties from $\alpha_s$ and the quark masses themselves
are correlated in a complicated way (i) across the moments, (ii) across the two
quark flavors, (iii) between the theoretical moments and the continuum 
contribution, and (iv) with each other.  In practice, all this is accounted 
for by performing fits to the moments.  Heavy quark radiation by light 
quarks~\cite{Portoles:2002rt} is not resonating and problems associated with
singlet contributions~\cite{Portoles:2002rt,Groote:2001py} appear only at 
${\cal O} (\alpha_s^3)$, so these issues should not introduce further 
uncertainties into our analysis.  We will present our final results 
after discussing the $\tau$ lifetime.

For our analysis of the $\tau$ mean lifetime,
\be
   \tau_\tau = {\hbar\over \Gamma_\tau} = \hbar {1 - {\cal B}_S
   \over \Gamma_\tau^e + \Gamma_\tau^\mu + \Gamma_\tau^{ud}} =
   290.96 \pm 0.59 \mbox{ fs},
\label{tautau}
\ee
we evaluate the partial widths into leptons, 
$\Gamma_\tau^e + \Gamma_\tau^\mu$, and hadrons with vanishing net strangeness, 
$\Gamma_\tau^{ud}$, theoretically.  The relative fraction of decays with 
$\Delta S = -1$, ${\cal B}_S = 0.0286 \pm 0.0009$~\cite{Hagiwara:2002}, is based 
on experimental data, since the value for the strange quark mass, 
$\hat{m}_s (m_\tau)$, is not
well known, and the PQCD expansion, $C^{D=2}_{QCD}$, proportional to $m_s^2$ 
converges poorly and cannot be trusted. $C^{D=2}_{QCD}$ also multiplies
the corresponding $m_{u,d}^2$ terms in $\Gamma_\tau^{ud}$, posing the same but 
numerically less important problem there. We solved it, by relating 
$C^{D=2}_{QCD}$ to the ratio $\Gamma_\tau^{us}|V_{ud}|^2/(\Gamma_\tau^{ud} 
|V_{us}|^2) = 0.896\pm 0.034$~\cite{Hagiwara:2002} (in which to linear
order all universal terms cancel), and find 
$C^{D=2}_{QCD} (m_s^2 - m_d^2) = m_\tau^2 (0.091 \pm 0.046)$.
We included one-loop electroweak ($S_{\rm EW}$)~\cite{Marciano:vm} and QED 
(phase space) corrections~\cite{Nir:1989rm}, quark condensate contributions, 
as well as $c$ quark effects in an expansion in 
$m_\tau^2/4 m_c^2$~\cite{Larin:1994va}. 
{\em E.g.},
$$ 
   \Gamma_\tau^{ud} = {G_F^2 m_\tau^5 |V_{ud}|^2\over 64\pi^3}S_{\rm EW}
   ( 1 + {3 m_\tau^2\over 5 M_W^2} ) \left[  F_{\rm QCD} +
   {\hat\alpha\over \pi} ({85\over 24} - {\pi^2\over 2}) \right.
$$    
\vspace{-12pt}                 
$$
   \left. - 0.09 {m_u^2 + m_d^2\over m_s^2 - m_d^2} -
   {f_{\pi^\pm}^2\over m_\tau^4} [m^2_{\pi^\pm} (8\pi^2 + 23 \alpha_s^2) 
   - 4 m^2_{K^\pm} \alpha_s^2] \right].
$$
$F_{\rm QCD} - 1$ is the massless QCD 
correction.  Due to effects governed by the $\beta$-function, the 3-loop PQCD 
expansion~\cite{Braaten:1991qm} shows slow convergence.  The series
can be reorganized~\cite{LeDiberder:1992te} into a well-behaved expansion with
coefficients, $d_i$~\cite{Chetyrkin:1996ia}, from the Adler $D$-function.  
The new expansion is not a power series. Rather, the $d_i$ multiply 
complicated functions, $A_i(\alpha_s)$~\cite{LeDiberder:1992te}, which we 
calculate numerically up to 4-loop order in the $\beta$ 
function~\cite{vanRitbergen:1997va}.

We computed the world average~(\ref{tautau}) by combining the direct value, 
$\tau_\tau = 290.6\pm 1.1$~fs~\cite{Hagiwara:2002}, with
$\tau_\tau ({\cal B}_e,{\cal B}_\mu) = 291.1 \pm 0.7$~fs derived from 
the leptonic branching ratios ${\cal B}_e = 0.1784(6)$ and 
${\cal B}_\mu = 0.1737(6)$~\cite{Hagiwara:2002} taking into account their 1\% 
correlation. The dominant theoretical error induced by the unknown 
coefficient $d_3 = 0 \pm 77$~\cite{Erler:1999ug} is itself strongly 
$\alpha_s$-dependent, is recalculated in each call within a fit, and induces 
an asymmetric $\alpha_s$ error.  

Other experimental uncertainties arise from~\cite{Hagiwara:2002}
$m_\tau = 1.77699(28)$~GeV, $|V_{ud}| =  0.97485(46)$, and 
${\cal B}_S$. Uncertainties from higher dimensional terms in 
the operator product expansion, OPE, are taken from Ref.~\cite{Barate:1998uf} 
and add up to $\Delta \tau_\tau ({\rm OPE}) = \pm 0.64$~fs.  We assume that 
an uncertainty of the same size is induced by possible OPE breaking 
effects\footnote{It is sometimes speculated that OPE breaking effects could
induce dangerous terms of ${\cal O}(\Lambda_{\rm QCD}^2/m_\tau^2)$.
The absence of numerically significant terms of that type is difficult to prove
with rigor. We stress, however, that the fits to OPE condensate terms of 
Ref.~\cite{Barate:1998uf} should have revealed their presence.}.
The unknown five-loop $\beta$-function coefficient, 
$\beta_4 = 0 \pm 579$~\cite{Erler:1999ug}, contributes mainly to the evolution 
of $\alpha_s (m_\tau)$ to $\alpha_s (M_Z)$ and less to the $A_i$. 
The subleading errors listed in this paragraph amount to $\pm 1.2$~fs. 
We find, $\alpha_s (m_\tau) = 0.356_{-0.021}^{+0.027}$ and 
$\alpha_s (M_Z) = 0.1221_{-0.0023}^{+0.0026}$, in excellent agreement with 
$\alpha_s (M_Z) = 0.1200 \pm 0.0028$ from $Z$-decays~\cite{Erler:sa} and most
other recent evaluations of $\tau_\tau$~\cite{Barate:1998uf,Raczka:1994ha}. 
Including $\tau_\tau$, and the $n=2$ and $n=6$ moments for the $c$ and $b$ 
quark, respectively, as constraints in a fit to all data~\cite{Erler:sa}
yields,
\be
\ba{l}
  \alpha_s (M_Z) = 0.1211_{-0.0017}^{+0.0018}, \vspace{4pt} \\
  \hat{m}_c (\hat{m}_c)  = 1.289^{+0.040}_{-0.045}~{\rm GeV}, \vspace{4pt} \\
  \hat{m}_b (\hat{m}_b)  = 4.207^{+0.030}_{-0.031}~{\rm GeV}.
\ea
\ee
These results reduce the error~\cite{Erler:1998sy} in $\alpha(M_Z)$ by 25\%.


\begin{acknowledgments} 
We would like to thank Jing Wang for collaboration in the early stages
of this work and for helpful discussions.
JE was supported by CONACYT (Mexico) contract 42026-F and by DGAPA-UNAM contract PAPIIT IN112902.
ML was supported in part by the Fund for Trans-Century Talents, CNSF-90103009 and CNSF-10047005.
\end{acknowledgments}


\begin{thebibliography}{99}

\bibitem{Langacker:1980js}
P.~Langacker, \pr{72}{185}(1981).

\bibitem{Hagiwara:2002}
K.~Hagiwara \etal, \prt{66}{010001}(2002).

\bibitem{Erler:sa}
J.~Erler and P.~Langacker, p.\ 98 in Ref.~\cite{Hagiwara:2002}.

\bibitem{Abreu:1997ey}
DELPHI: P.~Abreu \etal, \plb{418}{430}(1998).

\bibitem{Braaten:1991qm}
E.~Braaten, S.~Narison and A.~Pich, \npb{373}{581}(1992).

\bibitem{Novikov:et}
V.~A.~Novikov \etal, \pr{41}{1}(1978).

\bibitem{Shifman:bx}
M.~A.~Shifman, A.~I.~Vainshtein and V.~I.~Zakharov, \npb{147}{385}(1979)
and \ibid \con{147}{448}(1979).

\bibitem{Erler:1998sy}
J.~Erler, \prt{59}{054008}(1999).

\bibitem{Voloshin:1995sf}
M.~B.~Voloshin, \ijm{10}{2865}(1995);
K.~Melnikov and A.~Yelkhovsky, \prt{59}{114009}(1999);
A.~A.~Penin and A.~A.~Pivovarov, \npb{549}{217}(1999);
M.~Jamin and A.~Pich, \npp{74}{300}(1999);
M.~Beneke and A.~Signer, \plb{471}{233}(1999);
S.~Narison, \plb{520}{115}(2001);
A.~H.~Hoang, 
{\tt hep-ph/0008102}.

\bibitem{Eidemuller:2000rc}
C.~A.~Dominguez, G.~R.~Gluckman and N.~Paver, \plb{333}{184}(1994);
M.~Eidem\"uller and M.~Jamin, \plb{498}{203}(2001).
J.~Pe\~narrocha and K.~Schilcher, \plb{515}{291}(2001).

\bibitem{Kuhn:2001dm}
J.~H.~K\"uhn and M.~Steinhauser, \npb{619}{588}(2001).

\bibitem{Chetyrkin:1996cf}
K.~G.~Chetyrkin, J.~H.~K\"uhn and M.~Steinhauser, \npb{482}{213}(1996).

\bibitem{Chetyrkin:1997un}
K.~G.~Chetyrkin, B.~A.~Kniehl and M.~Steinhauser, \npb{510}{61}(1998).

\bibitem{Chetyrkin:1996ia}
K.~G.~Chetyrkin, J.~H.~K\"uhn and A.~Kwiatkowski, \pr{277}{189}(1996).

\bibitem{Bai:2001ct}
BES Collaboration: J.~Z.~Bai \etal, 
{\tt hep-ex/0102003}.

\bibitem{Chetyrkin:1997mb}
K.~G.~Chetyrkin, J.~H.~K\"uhn and M.~Steinhauser, \npb{505}{40}(1997).

\bibitem{Broadhurst:1994qj}
D.~J.~Broadhurst \etal, \plb{329}{103}(1994).

\bibitem{Erler:1999ug}
J.~Erler, 
{\tt hep-ph/0005084}.

\bibitem{Portoles:2002rt}
J.~Portoles and P.~D.~Ruiz-Femenia, 
{\tt hep-ph/0202114}.

\bibitem{Groote:2001py}
S.~Groote and A.~A.~Pivovarov, 
{\tt hep-ph/0103047}.

\bibitem{Marciano:vm}
W.~J.~Marciano and A.~Sirlin, \pru{61}{1815}(1988);
E.~Braaten and C.~S.~Li, \prt{42}{3888}(1990).

\bibitem{Nir:1989rm}
Y.~Nir, \plb{221}{184}(1989).

\bibitem{Larin:1994va}
S.~A.~Larin, T.~van Ritbergen and J.~A.~Vermaseren, \npb{438}{278}(1995).

\bibitem{LeDiberder:1992te}
A.~A.~Pivovarov, {\em Sov. J. Nucl. Phys.} {\bf 54}, 676 (1991) and
\zpc{53}{461}(1992);
F.~Le Diberder and A.~Pich, \plb{286}{147}(1992) and \ibid 
\con{289B}{165}(1992).

\bibitem{vanRitbergen:1997va}
T.~van Ritbergen, J.~A.~Vermaseren and S.~A.~Larin, \plb{400}{379}(1997).

\bibitem{Barate:1998uf}
ALEPH: R.~Barate \etal, \epc{4}{409}(1998).

\bibitem{Raczka:1994ha}
P.~A.~Raczka and A.~Szymacha, \zpc{70}{125}(1996);
J.~G.~K\"orner, F.~Krajewski and A.~A.~Pivovarov, \prt{63}{036001}(2001);
B.~V.~Geshkenbein, B.~L.~Ioffe and K.~N.~Zyablyuk, \prt{64}{093009}(2001).

\end{thebibliography}
\end{document}